\documentclass[letterpaper]{article} % DO NOT CHANGE THIS
\usepackage{aaai24}  % DO NOT CHANGE THIS
\usepackage{times}  % DO NOT CHANGE THIS
\usepackage{helvet}  % DO NOT CHANGE THIS
\usepackage{courier}  % DO NOT CHANGE THIS
\usepackage[hyphens]{url}  % DO NOT CHANGE THIS
\usepackage{graphicx} % DO NOT CHANGE THIS
\usepackage{natbib}  % DO NOT CHANGE THIS AND DO NOT ADD ANY OPTIONS TO IT
\usepackage{caption} % DO NOT CHANGE THIS AND DO NOT ADD ANY OPTIONS TO IT
\usepackage{algorithm}
\usepackage{algorithmic}

\usepackage{newfloat}
\usepackage{listings}
\DeclareCaptionStyle{ruled}{labelfont=normalfont,labelsep=colon,strut=off} % DO NOT CHANGE THIS
\floatstyle{ruled}
\newfloat{listing}{tb}{lst}{}
\floatname{listing}{Listing}
\usepackage{bbding}
\usepackage{xspace}
\usepackage{amsmath}
\usepackage{amsfonts}
\usepackage{graphicx}
\usepackage{nicematrix}
\usepackage{diagbox}
\usepackage{cleveref}

\newcommand{\method}{\textsc{NeuroImagen}\xspace}

\title{Seeing through the Brain: Image Reconstruction of Visual Perception\\from Human Brain Signals}

\author{Yu-Ting Lan$^1$\footnote{Work done during Yuting's intern at Microsoft Research Asia, correspondence to Kan Ren, Yansen Wang and Weilong Zheng.}, Kan Ren$^2$, Yansen Wang$^2$, Wei-Long Zheng$^1$,\\Dongsheng Li$^2$, Bao-Liang Lu$^1$, Lili Qiu$^2$
}

\affiliations{
    \textsuperscript{\rm 1}Shanghai Jiao Tong University, \textsuperscript{\rm 2}Microsoft Research\\
    \{kanren, yansenwang\}@microsoft.com, weilong@sjtu.edu.cn
}

\nocopyright
\usepackage{bibentry}
\begin{document}

\maketitle

\begin{abstract}
\textit{Seeing is believing}, however, the underlying mechanism of how human visual perceptions are intertwined with our cognitions is still a mystery. Thanks to the recent advances in both neuroscience and artificial intelligence, we have been able to record the visually evoked brain activities and mimic the visual perception ability through computational approaches.
In this paper, we pay attention to visual stimuli reconstruction by reconstructing the observed images based on portably accessible brain signals, i.e., electroencephalography (EEG) data.
Since EEG signals are dynamic in the time-series format and are notorious to be noisy, processing and extracting useful information requires more dedicated efforts; 
In this paper, we propose a comprehensive pipeline, named \method, for reconstructing visual stimuli images from EEG signals. 
Specifically, we incorporate a novel multi-level perceptual information decoding to draw multi-grained outputs from the given EEG data.
A latent diffusion model will then leverage the extracted information to reconstruct the high-resolution visual stimuli images.
The experimental results have illustrated the effectiveness of image reconstruction and superior quantitative performance of our proposed method.
\end{abstract}

\section{Introduction}
Understanding cortical responses to human visual perception has emerged a research hotspot, which can significantly motivate the development of computational cognitive system with the knowledge of neuroscience \cite{palazzo2020decoding}.
Along with the rapid development of physiological techniques such as functional magnetic resonance imaging (fMRI) or electroencephalograph (EEG), it becomes possible to record the visually-evoked human brain activities for further analysis.
Thus, the research community put the attention onto these 
complicated
brain signal data and try to reconstruct the stimuli contents used for evoking human subjects in the experiments, for understanding and simulating the human visual perception.

One of the mainstream attempts to study the human visual perception is to reconstruct
the seen contents such as images \cite{takagi2023high} or videos \cite{chen2023cinematic} used to evoke the human subjective in the stimuli experiments, via computational approaches such as deep neural networks.
These works are mainly based on fMRI data \cite{allen2022massive}, while collecting these imaging data requires expensive devices and lacks of convenience for practical usage.
In contrast, EEG has provided a more expedient solution to record and analyze brain signals, yet few works are learning visual perception upon these brain signal data.
EEG data are commonly time-series electrophysiological signals recorded via electrodes placed upon the human scalp, while the subjects are watching some stimuli contents such as images which have also been temporally aligned to the recorded signals in the data.

Though more convenient, reconstruction of visual stimuli from EEG signals are more challenging than that from fMRI data.
First, EEG signals are in time-series data format, which is temporal sequence and quite different to the static 2D/3D images, leading to the challenge of the matching stimuli to the corresponding brain signal pieces.
Second, the effects of electrode misplacement or body motion result in severe artifacts in the data with quite low signal-to-noise ratio (SNR), which have largely influenced the modeling and understanding of the brain activities.
Simply mapping the EEG input to the pixel domain to recover the visual stimuli is of low quality.
The existing works tackling image reconstruction from EEG either traditional generative models from scratch \cite{kavasidis2017brain2image} and fine-tuning large generative models \cite{bai2023dreamdiffusion}, which are less efficient; or just retrieving similar images from the data pool \cite{ye2022see}.
They fail to capture semantic information or reconstruct high-resolution outputs.

In this work, we propose a comprehensive pipeline for Neural Image generation, namely \method, from human brain signals.
To tackle with aforementioned challenges in this task, we incorporate a multi-level semantics extraction module which decodes different semantic information from the input signal with various granularity.
Specifically, the extracted information contains sample-level semantics which is easy to decode, and pixel-level semantics such as the saliency map of silhouette information that tends to more decoding difficulties. 
The multi-level outputs are further fed into the pretrained diffusion models with the control of the generation semantics.
Through this way, our method can flexibly handle the semantic information extraction and decoding problem at different granularity and difficulties, which can subsequently facilitate the generation via effectively controlling the fixed downstream diffusion model at different levels.

We evaluate our methods with the traditional image reconstruction solutions on EEG data.
The results demonstrate the superiority of our \method over the compared methods in both quantitative and qualitative results on the EEG-image dataset. The proposed multi-level semantics extraction at different granularity can largely increase the structure similarity and semantic accuracy of reconstructed images with the observed visual stimuli. 

\begin{figure*}
    \centering
    \includegraphics[width=\linewidth]{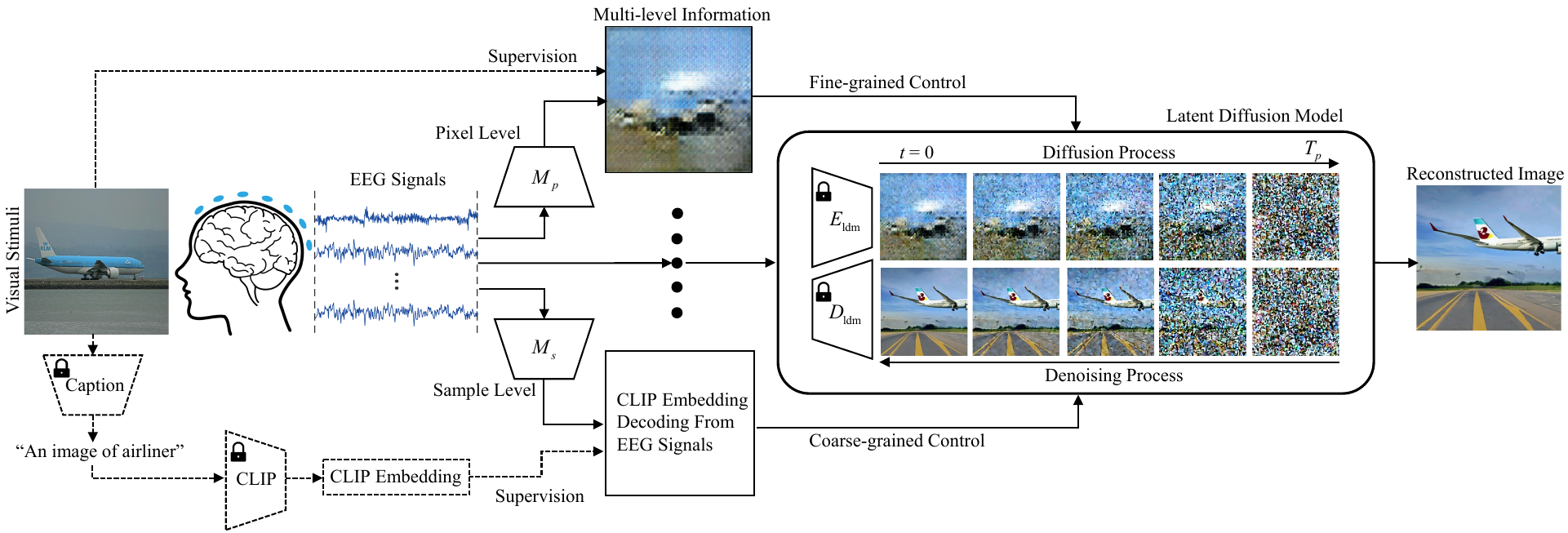}
    \caption{Overview of our \method. All the modules with dotted lines, \emph{i.e.} pixel-level supervision and sample-level supervision, are only used during the \emph{training} phase.
    and would be removed during the \emph{inference phase}.
    }
    \label{fig:model}
\end{figure*}

\section{Related Work}
\subsection{Diffusion Models}
Recently, diffusion models have emerged as state-of-the-art approaches in the field of generative models for several tasks, including image synthesis, video generation, and molecule design \cite{yang2022diffusion,song2020denoising,dhariwal2021diffusion}. 
A denoising diffusion probabilistic model (DDPM) \cite{ho2020denoising,sohl2015deep} is a parameterized bi-directional Markov chain using variational inference to produce sample matching after a finite time. 
The forward diffusion process is typically designed with the goal to transform any data distribution into a simple prior distribution (\emph{e.g.}, an isotropic Gaussian), and the reverse denoising diffusion process reverses the former by learning transition kernels parameterized by deep neural networks, such as U-Net \cite{ronneberger2015u}.
However, DDPM operates and undergoes evaluation and optimization in pixel space, leading to slower inference speeds and higher training costs.
To address these limitations, \citeauthor{rombach2022high}\shortcite{rombach2022high} introduced the concept of latent diffusion models (LDMs). 
In LDMs, diffusion models are applied within the latent space of the powerful pretrained autoencoders. 
This approach proves to be an effective generative model, accompanied by a separate compression stage that selectively eliminates imperceptible details. 
By operating in the latent space, LDMs overcome the drawbacks of pixel space evaluation, enabling faster inference and reducing training costs. 

\subsection{Image Decoding from fMRI}
The most recent works reconstructing the stimuli contents from the brain activities are mainly focused on fMRI data.
fMRI, as the measurement of the brain's blood-oxygen-level-dependent (BOLD) signals, has seen substantial advancements in brain signal decoding. 
The conventional visual decoding methods in fMRI usually rely on training deep generative neural networks, such as generative adversarial networks (GAN) and variational autoencoders (VAE) with paired brain-image data \cite{shen2019end,beliy2019voxels}. 
However, the decoding performance of these conventional methods is usually limited, and they struggle to produce visual contents with high resolution, because training a deep generative model from scratch is in general challenging and the dataset of brain signals is relatively small and noisy. 
Thus, recent research attempts to directly map brain signals into carefully pretrained latent spaces and finetune large-scale pretrained models to generate diverse and high-resolution images. 
\citeauthor{takagi2023high} map the brain activities to latent space and convert them to natural images using LDM. 
MinD-Vis \cite{chen2023seeing} integrates mask brain modelings and LDM to generate more plausible images with preserved semantic information.  
\citeauthor{zeng2023controllable} integrate silhouette information from brain signals with a controllable diffusion model to reconstruct high-quality images consistent with original visual stimuli. 
These methods generate more plausible and semantically meaningful images.

\subsection{Image Decoding from EEG Signals}
EEG is more portable but has relatively lower spatial resolution and suffers from larger noise, compared to fMRI, which makes decoding visual experience from brain signals a challenging problem. 
Brain2Image \cite{kavasidis2017brain2image} implements long short-term memory (LSTM) stacked with GAN and VAE techniques to generate seen images of ImageNet \cite{krizhevsky2012imagenet} from EEG signals \cite{kavasidis2017brain2image}. 
Neurovison \cite{khare2022neurovision} proposes conditional progressive growing of GAN (CProGAN) to develop perceived images and showed a higher inception score. \citeauthor{ye2022see} focuses on cross-modal alignment and retrieves images at the instance level, ensuring distinguishable model output for EEG signals.
We also note that there is a parallel work DreamDiffusion \cite{bai2023dreamdiffusion} to ours, which finetunes the diffusion model with an auxiliary task for aligning the EEG data and Image CLIP embeddings. 
However, the end-to-end training framework of DreamDiffusion struggles to effectively decode and utilize multi-level semantic information from EEG signals, which limits its ability to handle inherent noise characteristics. 
In addition, DreamDiffusion requires fine-tuning the diffusion models, which poses practical challenges and limitations in terms of scalability and efficiency.

\section{Methodology}
In this section, we design our method, \method, to extract multi-level semantics from EEG signals and subsequently integrate them into a pretrained diffusion model to reconstruct the observed visual stimuli from EEG signals.

We briefly introduce the intuition of multi-level semantics extraction in our \method. 
EEG signals are non-stationary time-series signals and are easy to disturb by artifacts like body motion, resulting in the low SNR of the signals. 
To tackle this challenge, we decode different semantic information with various granularity. 
Specifically, the pixel-level semantics such as the saliency map of silhouette information preserve fine-grained color, position, and shape details of the observed stimuli. 
The sample-level semantics provides a coarse-grained description of the visual stimuli, such the concept or category of the visual content. 
These designs exhibit the capacity to effectively manage the challenges posed by noisy time-series EEG data, consequently facilitating the reconstruction of high-quality visual stimuli.

In the following, we first formulate the problem and give an overview of \method. 
Then, we describe the multi-level semantics extractions of the \method, including pixel-level semantics and sample-level semantics with the corresponding training details of the decoding procedure. 
Finally, we detail the image reconstruction procedure of \method, which integrates the coarse-grained and fine-grained semantics with a pretrained latent diffusion model to reconstruct the observed visual stimuli from EEG signals. 

\subsection{Problem Statement}
In this section, we formulate the problem and give an overview of \method. 
Let the paired $\{(\text{EEG}, \text{image})\}$ dataset as $\Omega=\left\{\left(x_{i}, y_i\right)\right\}_{i=1}^n$, where $y_i \in \mathcal{R}^{H \times W \times 3}$ is the visual stimuli image to evoke the brain activities and $x_i \in \mathcal{R} ^{C \times T}$ represents the recorded corresponding EEG signals.
Here, $C$ is the channel number of EEG sensors and $T$ is the temporal length of a sequence associated with the observed image. 
The general objective of this research is to reconstruct an image $y$ using the corresponding EEG signals $x$, with a focus on achieving a high degree of similarity to the observed visual stimuli.

\subsection{Multi-level Semantics Extraction Framework}

\Cref{fig:model} illustrates the architecture of \method.
In our approach, we extract multi-level semantics, represented as $\{M_1(x), M_2(x), \cdots, M_n(x)\}$, which capture various granularity ranges from coarse-grained to fine-grained information from EEG signals corresponding to visual stimuli. The coarse-grained semantics serves as a high-level overview, facilitating a quick understanding of primary attributes and categories of the visual stimuli. On the other hand, fine-grained semantics offers more detailed information, such as localized features, subtle variations, and small-scale patterns. The multi-level semantics are then fed into a high-quality image reconstructing module $F$ to reconstruct the visual stimuli 
$\hat{y} = F[M_1(x), M_2(x), \cdots, M_n(x)]$.
Specifically, we give two-level semantics as follows. Let $M_p$ and $M_s$ be the pixel-level semantic extractor and sample-level semantic extractor, respectively. 
Pixel-level semantics is defined as the saliency map of silhouette information $M_p(x) \in \mathcal{R}^{ H_p \times W_p \times 3}$. 
This step enables us to analyze the EEG signals in the pixel space and provide the rough structure information. 
Subsequently, we define the sample-level semantics as $M_s(x) \in \mathcal{R}^{L \times D_s}$, to provide coarse-grained information such as image category or text caption.

To fully utilize the two-level semantics, the high-quality image reconstructing module $F$ is a latent diffusion model. It begins with the saliency map $M_p(x)$ as the initial image and utilizes the sample-level semantics $M_s(x)$ to polish the saliency map and finally reconstruct $\hat{y} = F(M_p(x), M_s(x))$.

\subsection{Pixel-level Semantics Extraction}
In this section, we describe how we decode the pixel-level semantics, \emph{i.e.} the saliency map of silhouette information. The intuition of this pixel-level semantics extraction is to capture the color, position, and shape information of the observed visual stimuli, which is fine-grained and extremely difficult to reconstruct from the noisy EEG signal. However, as is shown in \Cref{main}, despite the low image resolution and limited semantic accuracy, such a saliency map successfully captures the rough structure information of visual stimuli 
from the noisy EEG signals. 
Specifically, our pixel-level semantics extractor $M_p$ consists of two components: 
(1) contrastive feature learning to obtain discriminative features of EEG signals and 
(2) the estimation of the saliency map of silhouette information based on the learned EEG features.

\subsubsection{Contrastive Feature Learning}
We use contrastive learning techniques to bring together the embeddings of EEG signals when people get similar visual stimulus, i.e. seeing images of the same class. 
The triplet loss \cite{schroff2015facenet} is utilized as
\begin{equation}
\begin{aligned}
    \mathcal{L}_{\text {triplet }}= \max ( 0, \beta+
    & \|f_\theta(x^a)-f_\theta(x^p)\|_2^2 \\ 
  - &
 \|f_\theta(x^a)-f_\theta(x^n)\|_2^2 ),
\label{triplet}
\end{aligned}
\end{equation}
where $f_\theta$ is the feature extraction function \cite{kavasidis2017brain2image} that maps EEG signals to a feature space. $x^a, x^p, x^n$ are the sampled anchor, positive, and negative EEG signal segments, respectively. The objective of \cref{triplet} is to minimize the distance between $x^a$ and $x^p$ with the same labels (the class of viewed visual stimuli) while maximizing the distance between $x^a$ and $x^n$ with different labels. To avoid the compression of data representations into a small cluster by the feature extraction network, a margin term $\beta$ is incorporated into the triplet loss.

\subsubsection{Estimation of Saliency Map} 
After we obtain the feature of EEG signal $f_\theta(x)$, we can now generate the saliency map of silhouette information from it and a random sampled latent $z\sim\mathcal{N}(0, 1)$, i.e.,
\begin{align*}
    M_p(x) = G(z, f_\theta(x)).
\end{align*}
$G$ denotes for the saliency map generator. In this paper, we use the generator from the Generative Adversarial Network(GAN) \cite{goodfellow2020generative} framework to generate the saliency map and the adversarial loss is defined as follows:
\begin{align}
    \mathcal{L}_{\text{adv}}^D=& \max (0,1-D(A(y), f_\theta(x)))+ \nonumber
    \\ & \max (0,1+D(A(M_p(x))),  f_\theta(x))),
    \label{eq:loss_d}\\
    \mathcal{L}_{\text{adv}}^G =& - D(A(M_p(x)), f_\theta(x)). \label{eq:loss_g}
\end{align}
In GAN, besides the generator $G$, a discriminator $D$ is introduced to distinguish between images from the generator and ground truth images $x$. It is optimized by minimizing the hinge loss \cite{lim2017geometric} defined in Equation \eqref{eq:loss_d}. $A$ is the differentiable augmentation function \cite{zhao2020differentiable}. To stabilize the adversarial training process and alleviate the problem of mode collapse, we add the mode seeking regularization \cite{mao2019mode}:
\begin{align}
\mathcal{L}_{\text{ms}} = - \left(\frac{d_x\left(G\left(z_1, f_\theta(x)\right), G\left(z_2, f_\theta(x)\right)\right)}{d_z\left(z_1, z_2\right)}\right),
\label{loss_ms}
\end{align}
where $d_*(\cdot)$ denotes the distance metric in image space $x$ or latent space $z$ and $z_1, z_2\sim \mathcal{N}(0, 1)$ are two different sampled latent vectors.

To enforce the accuracy of the generated saliency map from the visual stimuli, we use the observed image as supervision and incorporate the Structural Similarity Index Measure (SSIM) as well:
\begin{equation}
\mathcal{L}_{\text{SSIM}}=1 -  \frac{\left(2 \mu_{x} \mu_{M_p(x)}+C_1\right)\left(2 \sigma_{x} \sigma_{M_p(x)}+C_2\right)}{\left(\mu_{x}^2+\mu_{M_p(x)}^2+C_1\right)\left(\sigma_{x}^2+\sigma_{M_p(x)}^2+C_2\right)},
\label{SSIM-loss}
\end{equation}
where $\mu_x$, $\mu_{M_p(x)}$, $\sigma_{x}$, and $\sigma_{M_p(x)}$ represent the mean and standard values of the ground truth images and reconstructed saliency maps of the generator. $C_1$ and $C_2$ are constants to stabilize the calculation.

The final loss for the generator is the weighted sum of the losses:
\begin{equation}
\begin{aligned}
\mathcal{L}_G & = \alpha_1 \cdot \mathcal{L}_{\text{adv}}^G  + \alpha_2 \cdot  \mathcal{L}_{\text{ms}} +  \alpha_3 \cdot \mathcal{L_{\text{SSIM}}},
\end{aligned}
\end{equation}
and $\alpha_{i \in \{1,2,3\}}$ are hyperparameters to balance the loss terms.

\subsection{Sample-level Semantics Extraction}
As aforementioned, the EEG signals are notorious for their inherent noise, making it challenging to extract both precise and fine-grained information. Therefore, besides fine-grained pixel-level semantics, we also involve sample-level semantic
extraction methods to derive some coarse-grained information such as the category of the main objects of the image content.
These features have a relatively lower rank and are easier to be aligned.
Despite being less detailed, these features can still provide accurate coarse-grained information, which is meaningful to reconstruct the observed visual stimuli.

Specifically, the process $M_s$ will try to align the information decoded from the input EEG signals to some generated image captions, which are generated by some other additional annotation model such as Contrastive Language-Image Pretraining (CLIP) model \cite{radford2021learning}.
Below we detail the processes of image caption ground-truth generation and semantic decoding with alignment.

\begin{figure}[!ht]
    \centering
    \includegraphics[width=\linewidth]{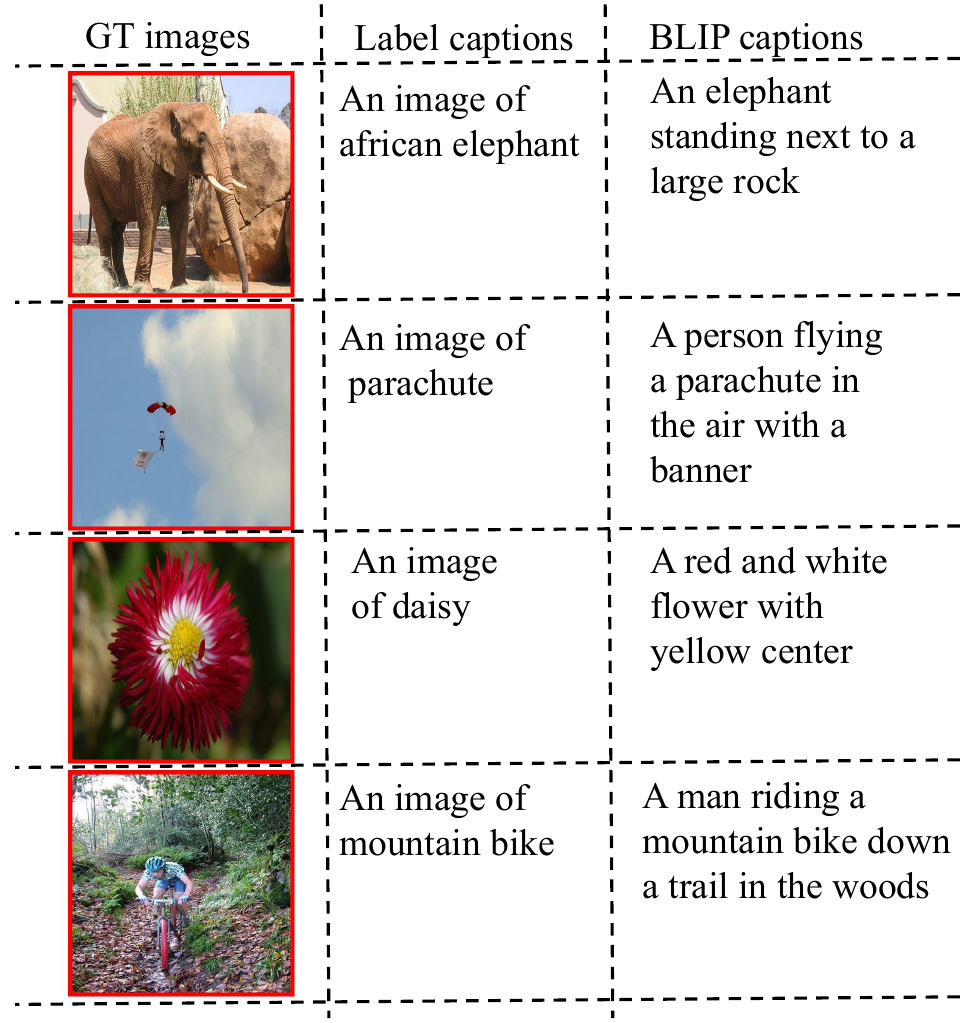}
    \caption{Examples of ground-truth images, label captions, and BLIP captions, respectively. 
    }
    \label{text_explanation}
\end{figure}

\subsubsection{Generation of Image Captions}
We propose two methods to generate the caption for each image to help supervise the decoding procedure of
semantic information from EEG signals.  
Since the observed images are from ImageNet dataset containing the class of the image,  we define a straightforward and heuristic method of label caption, which utilizes the class name of each image as the caption, as illustrated in the middle column of \Cref{text_explanation}.
The second method is that we use an image caption model BLIP \cite{li2023BLIP}, which is a generic and computation-efficient vision-language pretraining (VLP) model utilizing the pretrained vision model and large language models. 
We opt for the default parameter configuration of the BLIP model to caption our images. The examples are demonstrated in the right column of \Cref{text_explanation}. 
As can be seen, the label captions tend to focus predominantly on class-level information, and the BLIP-derived captions introduce further details on a per-image level. 

\begin{figure*}[!t]
    \centering
    \includegraphics[width=\linewidth]{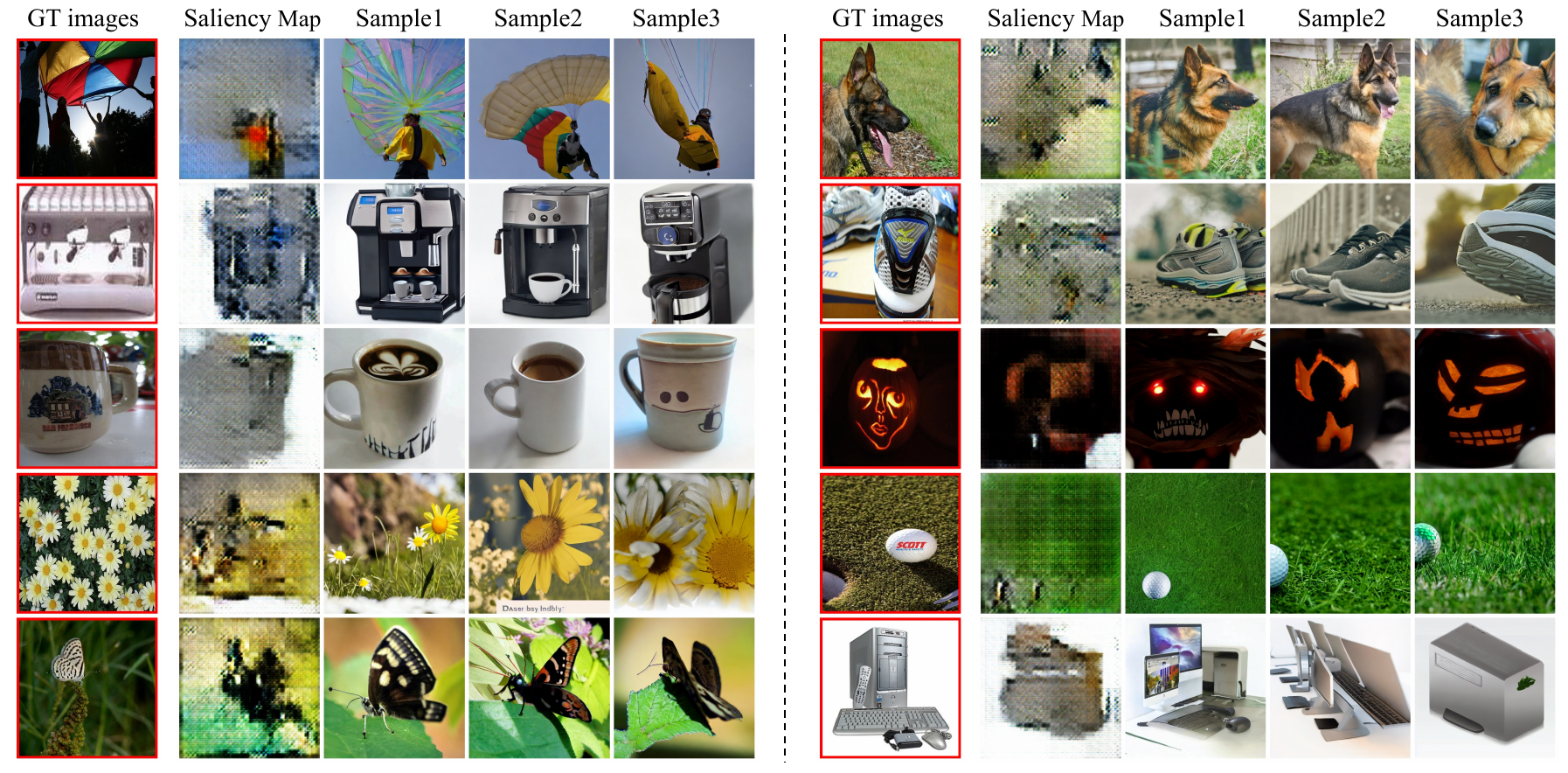}
    \caption{The main results of our \method. The images positioned on the left with red boxes represent the ground truth images. The second images from the left represent the pixel-level saliency map reconstructed from EEG signals. The three images on the right exhibit the three sampling results for the given saliency map under the guidance of sample-level semantics.}
    \label{main}
\end{figure*}
\begin{figure*}[!htt]
    \centering
    \includegraphics[width=\linewidth]{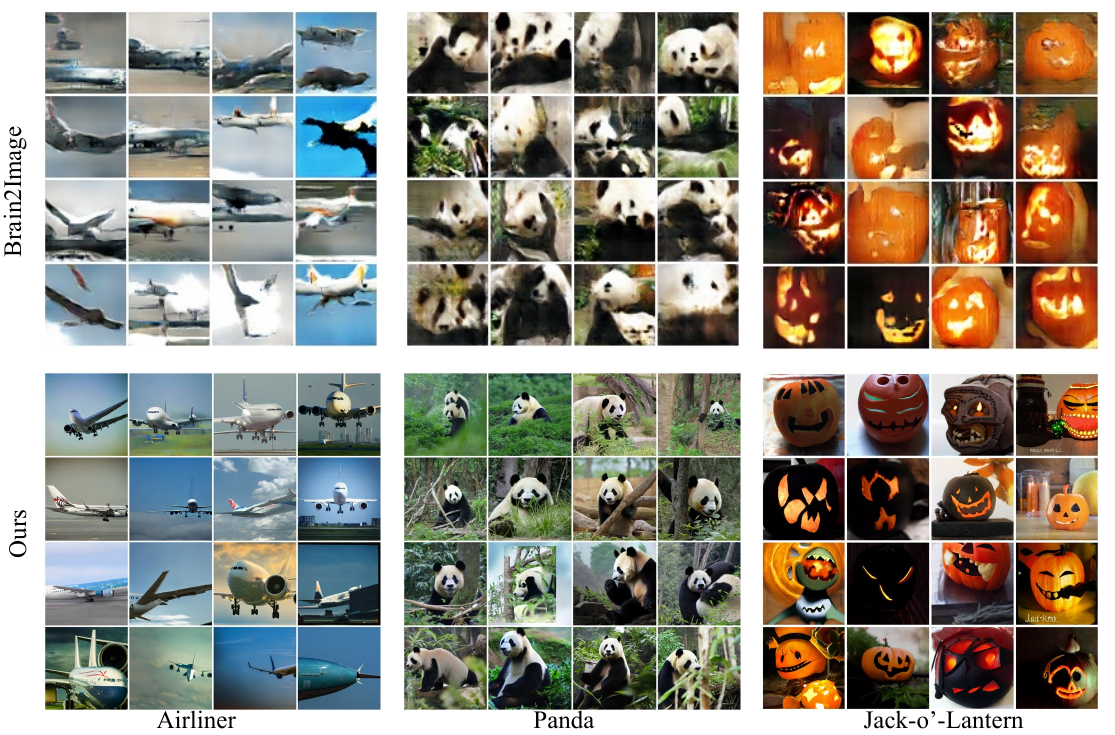}
    \caption{Comparison baseline Brain2Image \cite{kavasidis2017brain2image} and our proposed \method in three classes, namely 'Airliner', 'Panda', and 'Jack-o'-Lantern'. The first and second row depicts the results of Brain2Image and our \method, respectively.}
    \label{fig:compare2baselne}
\end{figure*}

\subsubsection{Predict the Text CLIP Embedding}
After the generation of the image caption ground-truth, the goal of the semantic decoding is to extract the information from the EEG signals to align the caption information.
Note that, this procedure is conducted in the latent space, where the latent embeddings have been processed from the CLIP model from the above generated captions.
Specifically, 
We extracted the CLIP embeddings $\hat{h}_{\text{clip}_{*}}$ from the generated captions and align the output $h_\text{clip}$ of EEG sample-level encoder with the loss function as
\begin{equation}
    \mathcal{L}_{\text{clip}} = || h_\text{clip} - \hat{h}_{\text{clip}_{*}} ||^{2}_2, 
\end{equation}
where $* \in \{B, L\}$ denotes the BLIP caption embedding or label caption embedding.

\subsection{Combining Multi-level EEG Semantics with Diffusion Model}

In this section, we present a comprehensive explanation of how multi-level semantics can be effectively integrated into a diffusion model for visual stimulus reconstruction. We utilize both pixel-level semantics, denoted as $M_p(x)$ (obtained using $G(z, f_{\theta}(x))$), and sample-level semantics, represented as $M_s(x)$ (obtained using $h_{clip}$), to exert various granularity control over the image reconstruction process. The reconstructed visual stimuli are defined as
$\hat{y} = F(M_p(x), M_s(x)) = F(G(f_\theta(x), h_{clip}))$.

Specifically, we used the latent diffusion model to perform image-to-image reconstructing with the guidance of conditional text prompt embeddings: 
(1) First, we reconstruct the pixel-level semantics $G(f_\theta{(x)})$ from EEG signals and resize it to the resolution of observed visual stimuli 
(2) $G(f_\theta{(x)})$ is then processed by the encoder $E_\text{ldm}$ of autoencoder from the latent diffusion model and added noise through the diffusion process. 
(3) Then, we integrate the sample-level semantics $h_\text{clip}$ as the cross-attention input of the U-Net to guide the denoising process. 
(4) We project the output of the denoising process to image space with $D_{\text{ldm}}$ and finally reconstruct the high-quality image $\hat{y}$.

\section{Experiments}
\subsection{Dataset}
The effectiveness of our proposed methodology is validated using the EEG-image dataset \cite{spampinato2017deep}. This dataset is publicly accessible and consists of EEG data gathered from six subjects. The data was collected by presenting visual stimuli to the subjects, incorporating 50 images from 40 distinct categories within the ImageNet dataset \cite{krizhevsky2012imagenet}. 
Each set of stimuli was displayed in 25-second intervals, separated by a 10-second blackout period intended to reset the visual pathway. This process resulted in totally 2000 images, with each experiment lasting 1,400 seconds (approximately 23 minutes and 20 seconds). 
The EEG-image dataset encompasses a diverse range of image classes, including animals (such as pandas), and objects (such as airlines).

Following the common data split strategy \cite{kavasidis2017brain2image}, we divide the pre-processed raw EEG signals and their corresponding images into training, validation, and testing sets, with corresponding proportions of 80\% (1,600 images), 10\% (200 images), and 10\% (200 images) and build one model for all subjects. The dataset is split by images, ensuring the EEG signals of all subjects in response to a single image are not spread over splits.

\subsection{Evaluation Metrics}
\subsubsection{$N$-way Top-$k$ Classification Accuracy (ACC)}
Following \cite{chen2023seeing}, we evaluate the semantic correctness of our reconstructed images by employing the $N$-way top-$k$ classification accuracy. Specifically, the ground truth image $y$ and reconstructed image $\hat{y}$ are fed into a pretrained ImageNet1k classifier \cite{dosovitskiy2020image}, which determines whether $y$ and $\hat{y}$ belong to the same class. Then we check for the reconstructed image if the top-$k$ classification in $N$ selected classes matches the class of ground-truth image. Importantly, this evaluation metric eliminates the need for pre-defined labels for the images and serves as an indicator of the semantic consistency between the ground truth and reconstructed images. In this paper, we select 50-way top-1 accuracy as the evaluation metric.

\subsubsection{Inception Score (IS)}  IS, introduced by \cite{salimans2016improved}, is commonly employed to evaluate the quality and diversity of reconstructed images in generative models.  To compute the IS, a pretrained Inception-v3 classifier \cite{szegedy2016rethinking} is utilized to calculate the class probabilities for the reconstructed images. We use IS to give a quantitative comparison between our method and baselines.

 \subsubsection{Structural Similarity Index Measure (SSIM)} SSIM offers a comprehensive and perceptually relevant metric for image quality evaluation. SSIM is computed over multiple windows of the ground truth image and the corresponding reconstructed image in luminance, contrast, and structure components, respectively.
\begin{figure*}[!htt]
    \centering
    \includegraphics[width=\linewidth]{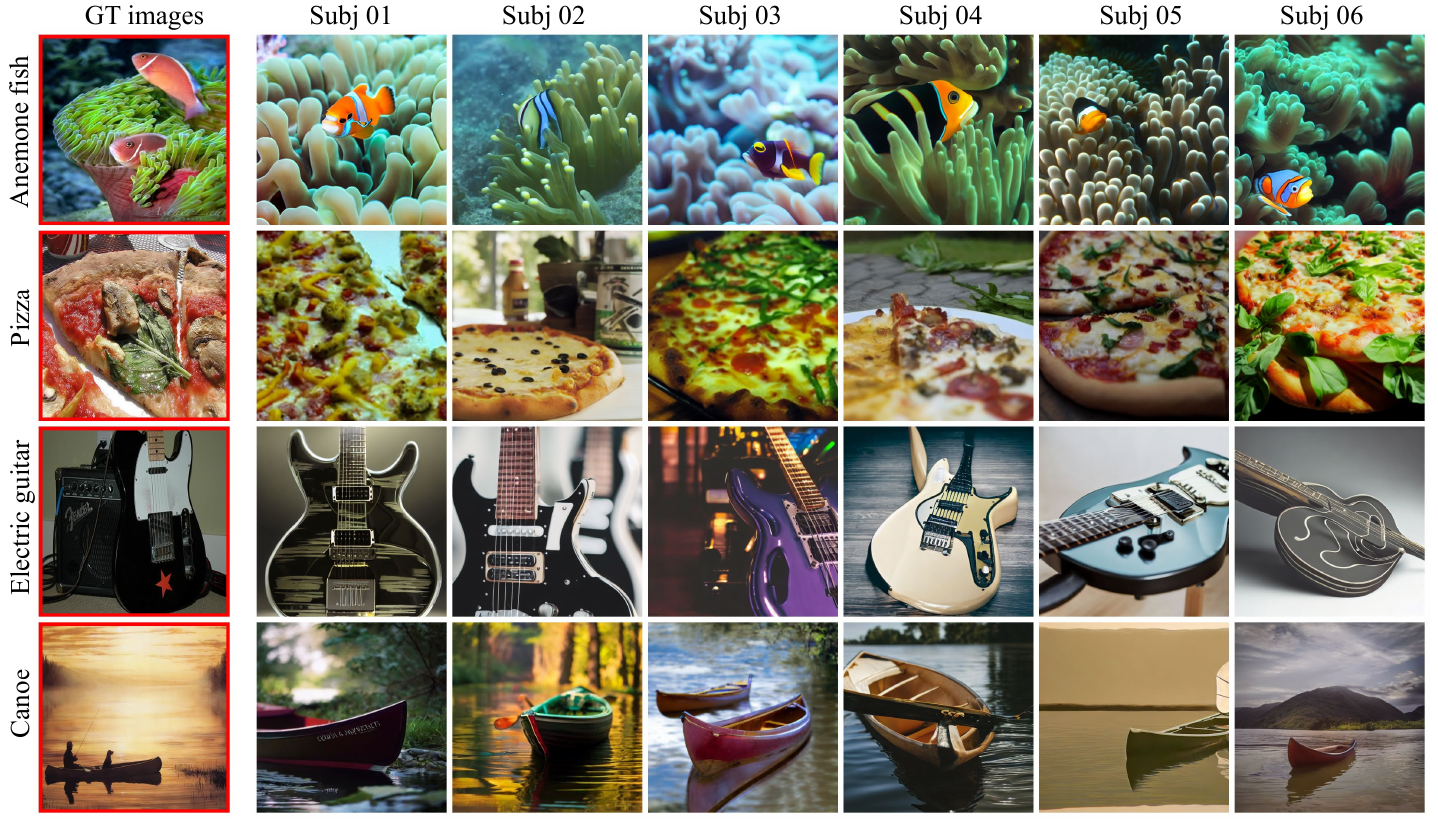}
    \caption{Comparison of reconstructed images on different subjects. The images on the left with red boxes represent the ground truth images. The other six images represent the reconstructed images of different subjects. The shown classes include fish, pizza, guitar, and canoe.}
    \label{cross_subject}
\end{figure*}

\section{Results}
\subsection{Experiment Results on the ImageNet Dataset}
The main results are illustrated in \Cref{main}.  The images positioned on the left with red boxes represent the ground truth images. The second images from the left represent the saliency map reconstructed from EEG signals. The three images on the right exhibit the three sampling results for the given pixel-level saliency map with the guidance of sample-level semantics of EEG signals. Upon comparison with the ground truth images and the reconstructed saliency maps, we validate that our pixel-level semantics extraction from EEG signals successfully captures the color, positional, and shape information of viewed images, despite limited semantic accuracy.  Comparing the GT images and three reconstructed samples,  it is demonstrated that the latent diffusion model successfully polishes the decoded saliency map with coarse-grained but accurate guidance of sample-level semantics from EEG signals. The high-quality reconstructed images purely from brain signals are perceptually and semantically similar to the viewed images.

\begin{table}[ht]
    \centering
        \begin{tabular}{c|c|c|c}
         \hline
            Model & ACC (\%) & IS  & SSIM\\
        \hline
        \hline
        Brain2Image & \diagbox{}{} & 5.01 & \diagbox{}{}\\
        \hline
        NeuroVision  & \diagbox{}{} & 5.23 & \diagbox{}{}\\
        \hline
        \textbf{\method}  & \textbf{85.6} & \textbf{33.50} & \textbf{0.249}\\
        \hline
        \end{tabular}
    \caption{The quantitative results of our \method, Brain2Image \cite{kavasidis2017brain2image} and  NeuroVision \cite{khare2022neurovision} on EEG-image dataset.}
    \label{tab_results}
\end{table}
\subsection{Comparison with Baselines}
The quantitative results of \method and baselines are listed in \Cref{tab_results}.  We have introduced the IS reported in the relevant literature, to exemplify the quality of the reconstructed images.
The IS is calculated by encompassing all images reconstructed across all subjects and all classes within the test set.
As is demonstrated in \Cref{tab_results}, the IS of our \method is significantly higher than Brain2Image and NeuroVision. Furthermore, inspired by \cite{bai2023dreamdiffusion}, we provide a qualitative comparison with the baselines in \Cref{fig:compare2baselne}. As can be seen, the quality of the images reconstructed by our \method is markedly higher than those reconstructed by the Brain2Image. This observed enhancement serves to validate the effectiveness and superiority of our proposed methodology.

\begin{table}[ht]
    \centering
        \begin{tabular}{c|c|c|c}
         \hline
            Subject & ACC (\%) & IS & SSIM  \\
        \hline
        \hline  
        subj 01  & 83.84 & 32.64 & 0.254 \\
        \hline
        subj 02 & 84.26 & 32.33 & 0.247 \\
        \hline
        subj 03 & 86.66 & 32.93 & 0.251\\
        \hline
        subj 04 & 86.48 & 32.40 & 0.244\\
        \hline
        subj 05 & 87.62 & 32.97 & 0.250\\
        \hline 
        subj 06 & 85.25 & 31.76 & 0.245\\
        \hline
        \end{tabular}
    \caption{The quantitative results of different subjects.}
    \label{tab:subject_results}
\end{table}

\begin{table}[ht]
    \centering
    \begin{tabular}{c|c|c|c|c|c|c}
    \hline
        Model & $B$ & $L$ & $I$ & ACC(\%) & IS & SSIM \\
    \hline
    \hline
        1 &  \XSolidBrush &  \XSolidBrush & \Checkmark & 4.5 & 16.31 & 0.234\\
    \hline
        2 &  \XSolidBrush & \Checkmark & \XSolidBrush & 85.9 & 34.12 & 0.180\\
    \hline
        3 &  \Checkmark & \XSolidBrush & \XSolidBrush & 74.1 & 29.87 & 0.157\\
    \hline
        4 &  \Checkmark & \XSolidBrush & \Checkmark & 65.3 & 25.86 & 0.235\\
    \hline
        5 &   \XSolidBrush & \Checkmark & \Checkmark & 85.6 & 33.50 & 0.249\\
    \hline
    \end{tabular}
    \caption{Quantitative results of ablation studies. $B$ and $L$ represent the semantic decoding using BLIP caption and label caption from EEG signals, respectively. $I$ represents the perceptual information decoding from EEG signals.}
    \label{tab:ablations}
\end{table}

\subsection{Generation Consistency in Different Subjects}
Since EEG signals are subject-specific cognitive processes that differ significantly in different subjects. In this section, we validate the robustness and feasibility of \method across different individuals. 
As is illustrated in \Cref{cross_subject}. The quantitative metric of different subjects are stable, which proves the generalization ability of \method. The qualitative results are shown in \Cref{cross_subject}. It can be seen the sampling from different subjects are semantically similar to the ground truth images.

\subsection{Ablation Study}
We further conduct experiments on the EEG-image dataset to analyze the effectiveness of each module of our \method. We define  $B$ and $L$ as the sample-level semantics from EEG signals using BLIP caption as supervision or label caption as supervision.  We define $I$ as the pixel semantics from EEG signals. The effectiveness of different methods is verified by employing ACC, IS, and SSIM.

\subsubsection{Pixel-level Semantics}
To demonstrate the effectiveness of the pixel-level semantics from EEG signals, we conduct validation on models 2, 3, 4, and 5. By comparing 2 with 5 and 3 with 4, we find that using the pixel-level semantics, \emph{i.e.} the saliency map, can significantly increase the structure similarity of the reconstructed images and ground truth images. 

\subsubsection{Sample-level Semantics}
We investigate the module of sample-level semantics decoding from EEG signals on guiding the denoising process. Models 1, 4, and 5 represent the experimental results only using the saliency, both the saliency map and sample-level semantics with the supervision of BLIP caption, and both the saliency map and sample-level semantics with the supervision of label caption,
respectively. By comparing 1 with 4 and
1 with 5, the experimental results
demonstrate that the use of sample-level semantics significantly increases the semantic accuracy of reconstructed images. 

\subsubsection{BLIP Captions vs Label Captions}
We also compare the two caption supervision methods with models 2 with 3 and 4 with 5. The experimental results of the label caption in all metrics are superior to using BLIP caption.  We impute these results to that the EEG signals may only capture the class-level information. So the prediction of BLIP latent is inaccurate, which decreases the performance of diffusion models.

\section{Conclusion}
In this paper, we explore to understand the visually-evoked brain activity. Specifically, We proposed a framework, named \method, to reconstruct images of visual perceptions from EEG signals. The \method first generates multi-level semantic information, i.e., pixel-level saliency maps and sample-level textual descriptions from EEG signals, then use the diffusion model to combine the extracted semantics and obtain the high-resolution images. Both qualitative and quantitative experiments reveals the strong ability of the \method.

As a preliminary work in this area, we demonstrate the possibility of linking human visual perceptions with complicated EEG signals. We expect the findings can further motivate the field of artificial intelligence, cognitive science, and neuroscience to work together and reveal the mystery of our brains to proceed visual perception information.

\bibliography{main}

\end{document}